\documentclass[aps,prl,reprint,superscriptaddress,showpacs,amsmath,amssymb]{revtex4-2}
\usepackage{xr}
\usepackage{graphicx}
\usepackage[utf8]{inputenc}
\usepackage{float}
\usepackage{dcolumn}
\usepackage{bm}
\usepackage{lineno}
\usepackage{rotating}
\usepackage{url}
\usepackage{braket}
\usepackage{mathrsfs}
\usepackage{overpic}

\usepackage[colorlinks,citecolor=blue,linkcolor=blue,breaklinks=true]{hyperref}
\begin{document}

\title{Measuring Spin-Charge Separation by an Off-diagonal Dissipative Response}

\author{Liang Tong}
%\email{ychen@gscaep.ac.cn}
\affiliation{Graduate School of China Academy of Engineering Physics, Beijing 100193, China}

\author{Shi Chen}
%\email{ychen@gscaep.ac.cn}
\affiliation{Graduate School of China Academy of Engineering Physics, Beijing 100193, China}

\author{Yu Chen}
\email{ychen@gscaep.ac.cn}
\affiliation{Graduate School of China Academy of Engineering Physics, Beijing 100193, China}

\begin{abstract}
Fractionalization of symmetry — exemplified by spin-charge separation in the 1D Hubbard model and fractional charges in the fractional quantum Hall effect — is a typical strongly correlated phenomena in quantum many-body systems. Despite the success in measuring velocity differences, however, it is still quite challenging in probing emergent excitations' anomalous dimensions experimentally. We propose a \emph{off-diagonal dissipative response} protocol, leveraging dissipative response theory (DRT), to directly detect spin-charge separation. By selectively dissipating spin-$\downarrow$ particles and measuring the spin-$\uparrow$ response, we uncover a universal temporal signature: the off-diagonal response exhibits a  crossover from cubic-in-time (\(t^3\)) growth at short times to linear-in-time (\(t\)) decay at long times. Crucially, the coefficients \(\varkappa^s\) (short-time) and \(\varkappa^l\) (long-time) encode the distinct anomalous dimensions and velocities of spinons and holons, providing unambiguous evidence of fractionalization. This signal vanishes trivially without spin-charge separation. Our predictions, verified numerically via tDMRG, with microscopic parameters linking with Luttinger parameters by Bethe ansatz, establish off-diagonal dissipative response as a probe of quantum fractionalization in synthetic quantum matter.
\end{abstract}

\maketitle

A hallmark of strongly correlated systems is the \emph{fractionalization} of quantum numbers, where emergent excitations exhibit deconfined fractional charges or spin-charge decoupling absent in the original microscopic framework. Paradigmatic examples include spin-charge separation in the 1D Hubbard model \cite{LiebWu68} and fractional charge excitations in the fractional quantum Hall effect (FQHE)\cite{Tsui82,Laughlin83}. In the 1D Hubbard case, electrons fractionalize into spinons and holons ---distinct quasiparticles propagating with separate velocities and anomalous dimensions. Similarly, FQHE systems host elementary excitations with quantized fractional charges (e.g., $e/3$) \cite{Heiblum97}. Focusing on spin-charge separation, although spectroscopic measurements and other dynamical or indirect measurements identified the existence of spin-charge separation \cite{Shen96, Baer99, West05, Shen06, Schofield09, Bloch17, Bloch20,Science2022}, however, accessing the key information such as the \emph{anomalous dimensions} of these excitations remain chanllenging in conventional spectroscopic probes.

Conventional measurement schemes rely on linear response theory, which probes systems via coherent external drivens \cite{mahan2000}. However, the spectroscopy detection for systems without well-defined quasiparticles are technically difficult due to requirements in homogenous density, wide spectrum range and fine spectrum resolution. This limitation can be overcome by alternative measurement protocols \cite{Duan06,ZD09,Niu06} including the dissipative response theory (DRT), where external drives are replaced as dissipative probes \cite{CH20}. DRT extracts critical information --- including anomalous dimensions and generalized sum rules \cite{Zhang24} --- from early-time dissipative dynamics of observables and entropy \cite{Chen21,Chen25}. This approach has proven useful in studying dissipative Bose-Hubbard systems \cite{ENS20,ENS23} and 1D interacting bosons via dissipation engineering \cite{Cirac09, Zoller08, Mueller22}, and we extend it here to uncover spin-charge separation in the fermionic Hubbard chain.

In this Letter, inspired by DRT, we propose an Off-diagonal dissipation response theory (Od-DRT) framework as an alternative scheme for detecting spin-charge separation. Central to this framework is a \emph{dissipative analog} of Hall response: measuring spin-$\uparrow$ particle response under selective dissipation targeting spin-$\downarrow$ particles. Here we will demonstrate that this response directly encodes the spin-charge separation, with the information of distinct anomalous dimensions of spinons and holons. To summarize in short, with the help of bosonization analysis, we find that the off-diagonal response follows a $t^3$-to-$t$ crossover with a single transition time scale. The velocity differences and anomalous dimension of spin and charge excitations are encoded in the coeffcients in coefficients of early-time $t^3$ law and late time $t$ law, making Od-DRT a direct probe of spin-charge separation. These results are confirmed numerically via time-dependent densitymatrix-renormalization-group (tDMRG) method, where microscopic parameters are mapped into Luttinger parameters by Bethe ansatz.

\emph{Setup and Off-diagonal Dissipative Response} --- We study an one-dimensional Hubbard model subjected to single-particle loss of spin-$\downarrow$ particles. The full Hamiltonian is $\hat{H} = \hat{H}_0 +\hat{H}_{\rm E}+\hat{V}$, composed of the system $\hat{H}_{0}$, the environment $\hat{H}_{\rm E}$, and their coupling $\hat{V}$. The system Hamiltonian is the standard Hubbard model
\begin{equation}
\hat{H}_0=-J\sum_{\left< i j \right>,\sigma}\hat{c}_{i\sigma}^\dag \hat{c}_{j \sigma} + U\sum_{i} \hat{n}_{i\uparrow}\hat{n}_{i\downarrow} -\mu\sum_{i,\sigma}\hat{n}_{i \sigma}, \label{Hubbard}
\end{equation}
where $\left< i j \right>$ denotes summation over nearest-neighbor sites $i, j = 1,2, \dots, L$ on a chain of length $L$. Here, $\hat{c}_{i \sigma}$ annihilates a fermion on site $i$ with spin $\sigma = \uparrow, \downarrow$, and $\hat{n}_{i \sigma}=\hat{c}_{i \sigma}^\dag\hat{c}_{i \sigma}$. Parameters are the hopping amplitude $J$ (set to $J=1$ as the energy scale), the on-site interaction $U$, and the chemical potential $\mu$. The system–environment coupling that induces loss of spin-$\downarrow$ particles is given by $\hat{V} = g\sum_{i} ( \hat{c}_{i \downarrow}^\dag \hat{\xi}_{i} + \hat{\xi}_i^\dag \hat{c}_{i \downarrow} )$, where $\hat{\xi}_i$ is a noise operator in the environment. In the low-temperature Markovian limit, the noise operators $\hat{\xi}_i(t)$ in the Heisenberg picture satisfy $\langle \hat{\xi}_i(t)\rangle_E =0$ and $\langle \hat{\xi}_i(t)\hat{\xi}_j^\dagger(t')\rangle_E = D\delta_{ij}\delta(t-t')$ under the environmental ensemble average $\langle \cdot \rangle_E$, while all other second-order correlators vanish.

Inspired by DRT \cite{CH20}, our Od-DRT framework probes the spin-$\uparrow$ response under selective spin-$\downarrow$ dissipation. This is achieved by monitoring the operator $\hat{W}_\uparrow = \hat{c}_{i\uparrow}^\dag\hat{c}_{j\uparrow}$, whose expectation value directly reflects the real-space momentum distribution of the spin-$\uparrow$ particles. Under periodic boundary conditions, and upon defining the distance $\ell = (i-j)/2$, the response function within Od-DRT is given by
\begin{equation}\label{Od_DRT}
\delta{\cal W} (\ell,t)= \gamma \sum_{m=1}^L\int_0^t dt'\left< {\cal L}_{\hat{\mathcal{O}}_{m\downarrow}(t')} \hat{W}_\uparrow(t) \right>,
\end{equation}
where $\gamma = g^2 D/2$ indicates the dissipation strength, and the superoperator ${\cal L}$ defined as ${\cal L}_{\hat{\mathcal{O}}_{m\downarrow}(t')} \hat{W}_\uparrow (t) \equiv 2\hat{\mathcal{O}}_{m\downarrow}^\dag(t')\hat{W}_\uparrow (t) \hat{\mathcal{O}}_{m\downarrow}(t')-\{\hat{\mathcal{O}}_{m\downarrow}^\dag(t')\hat{\mathcal{O}}_{m\downarrow}(t'),\hat{W}_\uparrow (t) \}$. Because the loss selectively couples only to spin-$\downarrow$ particles, the corresponding jump operator is given by $\hat{\mathcal{O}}_{m\downarrow}=\hat{c}_{m\downarrow}$. The expectation value $\langle \cdot \rangle$ denotes an average over the initial thermal ensemble. This response can be experimentally probed through measurements of the short-time dissipative dynamics.

An analytical treatment of the dissipative response in Eq.~(\ref{Od_DRT}) requires going beyond Wick’s theorem, which fails for mixed‑spin correlations and yields a vanishing result. We therefore employ bosonization --- a nonperturbative framework that captures interactions exactly and provides controlled access to the low‑energy physics of strongly correlated 1D  spin‑$1/2$ chains \cite{Giamarchi}. In the following, we use this technique to compute the relevant zero-temperature correlation functions, thereby obtaining a finite physical response.

\emph{Bosonization, Spin-charge separation in response}. ---  Bosonization provides the essential framework for describing one-dimensional interacting systems, where elementary excitations are not individual electrons but collective bosonic modes, such as particle–hole pairs and spin density waves \cite{Tomonaga1950,Luttinger1963}. In the 1D Hubbard model, although the fundamental degrees of freedom are electrons carrying both spin and charge, its low-energy excitations fractionalize into holons (charge \( \pm e \) without spin) and spinons (spin \( \pm \hbar/2 \) without charge) \cite{Haldane1981,PRL-64-2831}. This spin–charge separation is captured by introducing bosonic fields \( \phi_{a}(x) \) and their dual fields \( \theta_{a}(x) \) for the charge ($a = c$) and spin ($a = s$) channels, which satisfy the commutation relation:
\begin{equation}
[\phi_{a}(x), \partial_{y} \theta_{b}(y)] = i\pi \delta_{ab} \delta(x - y), \quad a,b \in \{c,s\}.
\end{equation}
The fermionic field operator can then be written as a vertex operator \cite{Delft1998,Giamarchi}:
\begin{eqnarray}\label{FFO}
  \hat{\psi}^{\dagger}_\sigma(x) &=&  \sqrt{\rho_0} \sum_{\varsigma = \pm 1 }  \hat{U}^{\dagger}_{\varsigma,\sigma} e^{-i \varsigma k_F x} \nonumber \\
  && \times \exp \left[ i \varsigma \phi_{c}(x) /\sqrt{2} \right] \exp \left[ - \theta_{c}(x) /\sqrt{2} \right]   \nonumber \\
  && \times \exp \left[ i \sigma \varsigma \phi_{s}(x) / \sqrt{2} \right] \exp \left[ - \sigma \theta_{s}(x)/\sqrt{2}\right],
\end{eqnarray}
where $\sigma = \pm 1$ denotes spin-$\downarrow, \uparrow$, $\rho_0$ is the average density per spin, $\varsigma = \pm 1$ labels right/left movers near $\pm k_F$, and the Klein factors $\hat{U}^{\dagger}_{\varsigma,\sigma}$ enforce fermionic statistics.

In the absence of external fields, the low‑energy universal behavior of the 1D Hubbard model is described by a two‑component Luttinger liquid (LL) \cite{Giamarchi,Hubbardbook,PRL-64-2831}. The effective bosonized Hamiltonian reads
\begin{equation} \label{Hubbard_B}
\hat{H}_0^{\rm B} = \frac{1}{2\pi} \sum_{a=c,s} \int dx \left[ u_a K_a (\partial_x \theta_a)^2 + \frac{u_a}{K_a} (\partial_x \phi_a)^2 \right], 
\end{equation}
where the charge and spin velocities $u_{c,s}$ and Luttinger parameters $K_{c,s}$ depend on the electron density $n = 2\rho_0$  and the interaction $U$. We focus on repulsive interactions ($U > 0$). In this regime, the SU(2) spin symmetry enforces $K_s = 1$, whereas the charge sector is characterized by $K_c < 1$. For systems that deviate from half-filling, umklapp scattering in the charge sector is irrelevant. Spin–charge separation manifests in two essential features: the collective charge and spin modes propagate at generically different velocities, $u_c \ne u_s$, and their correlation functions decay with distinct power laws. The asymptotic decay is governed by the anomalous dimensions
\begin{equation}
    \eta_{a} = \left(K_a + K_a^{-1} - 2 \right)/8, \quad a = c,s
\end{equation}
which depend separately on the Luttinger parameters of each sector. 

Implementing bosonization in Od-DRT requires first taking the continuous limit of the response function. The dissipative jump operator is taken as $\hat{\cal O}_{m\downarrow} \rightarrow \hat{\psi}_{\downarrow}(z)$ with spatial coordinate $z$, while $\hat{W}_{\uparrow} \rightarrow \hat{\psi}^\dagger_\uparrow (x) \hat{\psi}_\uparrow(y)$. By employing the bosonization technique, the characteristic structure of the response at the zero-temperature limit was obtained:
\begin{equation} \label{Od_DRT_B}
\delta{\cal W}(\ell,t) = \gamma n^2 \left( \frac{\alpha^2}{\alpha^2 + 4\ell^2}\right)^{\eta_c + \eta_s + \frac{1}{2}} \!\!\!{\cal F}(\ell) {\cal I}(\ell,t),
\end{equation}
where $\ell = (x-y)/2$  is the spatial separation, and the essential features are:
(i) a power-law spatial decay controlled by the exponents $\eta_c + \eta_s$;
(ii) an oscillatory factor ${\cal F}(\ell) = \cos\left[2k_F \ell\!-\!\tan^{-1}\!\left(2\ell/\alpha\right)\right]$;
(iii) a nontrivial dynamical kernel ${\cal I}(\ell,t)$ given by a time-dependent double integral that encapsulates the coupling of spin and charge modes:
\begin{eqnarray} \label{Od_DRT_Int}
{\cal I}(\ell,t)\!&=&\!\!\int_0^t \!\!d\tau \!\!
\int_{-\infty}^{\infty} \!\! d\xi\!\bigg[\prod_{\varsigma=\pm}\!\left(\frac{\mathfrak{F}_{+;1\!+\!2\eta_c,u_c}^\varsigma\mathfrak{F}_{-;2\eta_c,u_c}^\varsigma}{\mathfrak{F}_{+;1\!+\!2\eta_s,u_s}^\varsigma\mathfrak{F}_{-;2\eta_s,u_s}^\varsigma}\right)\!(\xi,\ell,\tau)\!\!\!\nonumber\\
&& \hspace{0.3cm} + \prod_{\varsigma=\pm}\!\left(\frac{\mathfrak{F}_{+;2\bar{\eta}_c,u_c}^\varsigma\mathfrak{F}_{-;2\bar{\eta}_c,u_c}^\varsigma}{\mathfrak{F}_{+;2\bar{\eta}_s,u_s}^\varsigma\mathfrak{F}_{-;2\bar{\eta}_s,u_s}^\varsigma}\right)\!(\xi,\ell,\tau)\!-\!2\bigg],\!
\end{eqnarray}
with $\bar{\eta}_{c,s} = \sqrt{\eta_{c,s} (\eta_{c,s} + 1/2)}$ and the chiral phase factors are defined as
\begin{equation} \label{CPF}
    \mathfrak{F}_{\pm;\lambda,u}^\varsigma(\xi,\ell,t) = \exp \left[ i\varsigma \lambda \tan^{-1}\left( \frac{u\tau\pm (\xi+\varsigma\ell)}{\alpha} \right) \right],
\end{equation}
where the index $\lambda \in \{ 1+2\eta_{c,s}, 2\eta_{c,s}, 2\bar{\eta}_{c,s} \}$ and $u=u_{c,s}$. Crucially, the full dissipative response $\delta{\cal W}(\ell,t)$ vanishes unless both anomalous dimensions $\eta_{c,s}$ and the velocity difference $\Delta u = u_c - u_s$ are non-zero, thereby providing a direct probe of spin-charge separation.

From the analytical form of Eq.~(\ref{Od_DRT_B}), one can extract the temporal asymptotics, which encode distinct physical information in the short- and long-time regimes. While the long-time limit recovers the steady-state Luttinger-liquid correlations, the short-time behavior is especially revealing: it directly captures the dynamical emergence of spin-charge separation, providing a clear probe of the independent propagation and distinct velocities of holons and spinons. To extract these limits, we adopt the piecewise-linear approximation: $\tan^{-1}(x)=-\frac{\pi}{2}\theta(\frac{\pi}{2}+x)+x\theta(\frac{\pi^2}{4}-x^2)+\frac{\pi}{2}\theta(x-\frac{\pi}{2})$, and evaluate the complex integral ${\cal I}(\ell,t)$ asymptotically. (A detailed derivation is given in the Supplemental Material.) In the short-time limit ($t < \frac{\pi \alpha}{2 u_c}$), the integral reduces to ${\cal I}(\ell, t) \approx {\cal I}^s t^3$, showing a pure cubic-in-time growth with no dependence on the spatial separation $\ell$. The coefficient ${\cal I}^s$ is given by
\begin{equation} \label{STI}
{\cal I}^s = -\frac{\pi}{3 \alpha}\! \bigg[ \!\left( 4\eta_c u_c \!-\! 4 \eta_s u_s  \!+\! \Delta u \right)^2 \!+ \!16(\bar{\eta}_c u_c \!-\! \bar{\eta}_s u_s  )^2 \! \bigg].
\end{equation} 
By contrast, in the long-time limit, the integral simplifies to ${\cal I}(\ell, t) \approx {\cal I}^l(\ell) t$, grows linearly in time.
Thus, we uncover that $\delta \mathcal{W}(\ell,t)$ exhibits a cubic-in-time dependence in the short-time limit and a linear time dependence at long-time limit:
\begin{equation}
    \delta {\cal W}(\ell,t) \approx
    \begin{cases} 
        \varkappa^s (\ell)t^3, & t \rightarrow 0, \\
        \varkappa^l(\ell) t, & t \rightarrow \infty.
    \end{cases}
\end{equation}
where the coefficients $\varkappa^s(\ell)$ and $\varkappa^l(\ell)$ follow directly from the asymptotic forms of ${\cal I}(\ell,t)$.

\begin{figure}[htpb]
\centering
\includegraphics[width=1.1\linewidth]{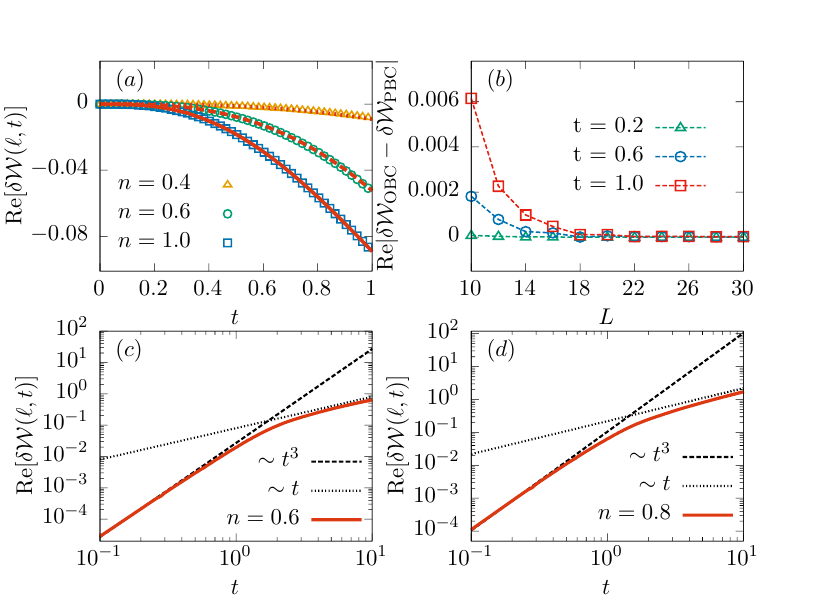}
\caption{(a) Real part of $\delta{\cal W}(\ell,t)$ versus time for $\ell = 1$ under three different densities $n$, comparing PBC with $L = 30$ (colored shapes) and OBC with $L = 50$ (lines). (b) Finite-size scaling analysis for the convergency of PBC and OBC results is performed at three time points along the evolution curve in (a) for $n = 1.0$. (c-d) Real part of $\delta{\cal W}(\ell,t)$ versus time at $\ell = 0.5$ on a log-log scale for $n = 0.6$ and 0.8, respectively, in a system of size $L=50$. Black dashed and dotted lines indicate reference slopes for $t^3$ and $t$, respectively. In all calculations, the interaction is fixed at $U = 5$.} \label{AFig1}
\end{figure}

\emph{tDMRG numerical experiment} --- To quantitatively validate our theoretical predictions, we performed large-scale numerical simulations using the zero-temperature tDMRG method \cite{PRL-93-040502,PRL-93-076401,SCHOLLWOCK201196}. This approach allows us to compute the exact time evolution of the equilibrium correlation function specified in the Od-DRT expression (Eq.~(\ref{Od_DRT})), effectively serving as a controlled numerical experiment that incorporates nonlinear Luttinger liquid effects, finite-size corrections, and additional contributions from umklapp scattering. Within the tDMRG framework, the many-body wave functions are represented as matrix product states (MPS) and evolved via the Trotter–Suzuki decomposition, with adaptive truncation of the Hilbert space guided by entanglement entropy optimization. Nonlinear fitting of the numerical data confirms that spin-charge separation is clearly extracted across a broad parameter range. In all simulations, the dissipation strength is fixed at $\gamma=1$.

While our bosonization analysis assumes periodic boundary conditions (PBC), the tDMRG simulations adopt open boundary conditions (OBC) for enhanced computational efficiency. In practice, we employ the time-evolving block decimation (TEBD) algorithm \cite{PRL-93-040502} under OBC, which we find to be substantially faster than the time-dependent variational principle (TDVP) method \cite{PRL-107-070601} under PBC for equivalent physical parameters. This advantage stems from TEBD's local update structure and the lower entanglement entropy in OBC, which allow larger time steps ($dt=0.005$) and smaller bond dimensions ($D=700 \sim 1000$) without loss of accuracy (truncation error $\sim 10^{-10}$).  In contrast, TDVP requires global updates and strict conservation enforcement along the entire ring, leading to substantially higher computational cost. Numerical benchmarks confirm that TEBD achieves accuracy comparable to TDVP while reducing runtime by at least an order of magnitude. 

In Fig.~\ref{AFig1}(a-b), we demonstrate the convergence between OBC and PBC results for sufficiently large system sizes. Here we only show the real part of $\delta{\cal W}(\ell,t)$ because the PBC restricts the imaginary part to be zero. The difference $|\delta \mathcal{W}_{\rm OBC} - \delta \mathcal{W}_{\rm PBC}|(\ell,t)$ in Fig.~\ref{AFig1}(b) decays monotonically with increasing system size $L$ and becomes negligible for $L > 30$, justifying the substitution of OBC observables for PBC values. For OBC simulations, the sites $i$ and $j$ are chosen near the lattice center $L/2$. For each fixed $\ell$, we average the values of $\delta \mathcal{W} (\ell,t)$ obtained by shifting the center $R = (i+j)/2$ by 5 sites, thereby minimizing finite-size boundary effects.

Figures \ref{AFig1}(c) and \ref{AFig1}(d) present the time evolution of the response functions for densities $n=0.6$ and $0.8$ on log–log scale, showing a transition from an initial $t^3$ rise to linear-in-$t$ growth. This $t^3$-to-$t$ law holds across a broad range of interaction strengths $U$ (refer to SM). A power-law fit to the early-time data gives an exponent $2.99 \pm 0.02$, in quantitative agreement with the predicted $t^3$ behavior. At later times, the response saturates to a linear asymptote, approached with increasing evolution time.

\begin{figure}[htpb]
\centering
\includegraphics[width=1.11\linewidth]{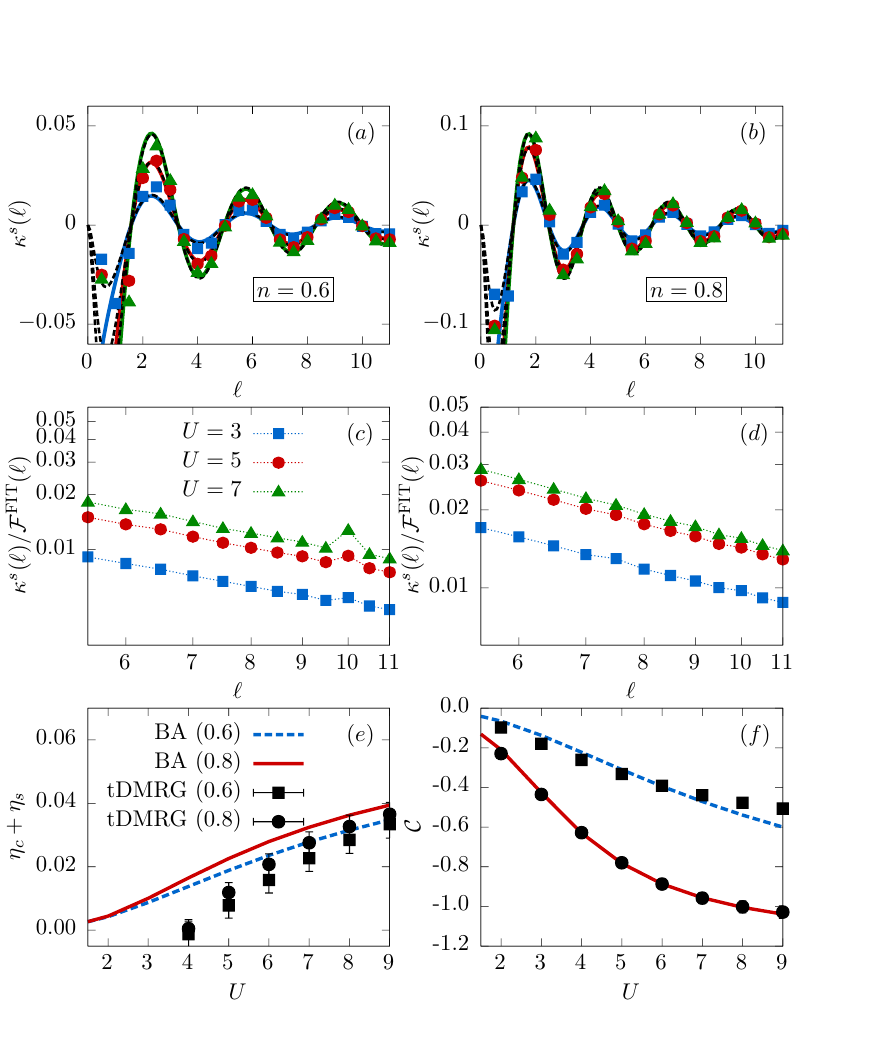}
\caption{(a,b) The short-time coefficient $\varkappa^{s}(\ell)$ obtained from tDMRG simulations (colored dots) for $U = 3,5,7$, compared with the analytical approximation (solid lines, Eq.~(\ref{STI})) and the exact numerical integration of Od-DRT (black dashed lines, Eq.~(\ref{Od_DRT_B})).
(c,d) Corresponding results after removing oscillations, revealing the power-law dependence in $\varkappa^s(\ell)$. Panels (c) and (d) are derived from the data in (a) and (b), respectively. (e) Extracted $\eta_{c} + \eta_{s}$ as a function of $U$, compared with Bethe ansatz results \cite{PRL-64-2831,Hubbardbook}. (f) The optimal overall amplitude as a function of $U$. One can see the curve is better fitted in $n=0.8$ then $n=0.6$.} \label{AFig2}
\end{figure}

In Fig.~\ref{AFig2}(a) and \ref{AFig2}(b), we present the $t^3$ coefficients $\varkappa^s(\ell)$ extracted from three interaction strengths $U=3,5,7$ under $n=0.6$ and 0.8. The amplitude of the oscillation decreases as the interaction weakens. For comparison, we present results from the analytical approximation (\ref{STI}) and the exact numerical integration (\ref{Od_DRT_B}). Both are rescaled via a constant factor that is solely a function of the filling $n$ and independent of $U$, to match the tDMRG data. Specifically, for $n=0.6$, the solid and dashed lines are multiplied by 0.195 and 0.44; for $n=0.8$, the corresponding factors are 0.245 and 0.59. Note that the $u_{c,s}$ and Luttinger parameters $K_{c,s}$ in the expressions (\ref{Od_DRT_B}) and (\ref{STI}) are determined based on the Bethe ansatz solutions \cite{LiebWu68,PRL-64-2831,Hubbardbook}. 

To directly extract the exponent $\eta_{c} + \eta_{s}$, we fit the numerically obtained $\varkappa^s(\ell)$ to the asymptotic Luttinger-liquid form (see SM for fitting details). From these fits we determine the oscillatory factor $\mathcal{F}^{\mathrm{FIT}}(\ell)$. As shown in Figs.~\ref{AFig2}(c) and \ref{AFig2}(d), the quantity $\varkappa^s(\ell)/\mathcal{F}^{\mathrm{FIT}}(\ell)$ exhibits a clean power-law decay in $\ell$. The corresponding exponents, denoted $(\eta_c + \eta_s)^{ \mathrm{FIT}}$, are plotted as a function of interaction strength $U$ in Fig.~\ref{AFig2}(e) and compared with the exact Bethe-ansatz results $(\eta_c + \eta_s)^{\mathrm{BA}}$. We emphasize that our fitting procedure does not presuppose the Luttinger-liquid framework. the agreement for moderate and strong $U$ thus provides a direct measurement of the anomalous dimensions. While the Bethe ansatz yields $\eta_s^{\rm BA}=0$ for the Hubbard model, we do not impose this constraint, and our measurement indeed returns $(\eta_c + \eta_s)^{ \mathrm{FIT}} \approx \eta_c^{\rm BA}$. Deviations between $(\eta_c + \eta_s)^{ \mathrm{FIT}}$ and $(\eta_c + \eta_s)^{ \mathrm{BA}}$ become noticeable at weaker interactions. This systematic discrepancy is consistent with the limitations of the linear Luttinger-liquid description, where non-linear dispersion effects become more pronounced in the regime of weak interactions and low density. The Od-DRT can therefore serve as a decisive probe of Fermi-edge singularity physics beyond the linear Luttinger model. The overall amplitude $\mathcal{C}$ extracted from the fits is compared with theory in Fig. \ref{AFig2}(f). The generally good agreement, especially for higher density, further validates the theoretical framework. The further theoretical analysis and numerical efforts will be left for future studies.

\emph{Summary and Outlook}. --- We propose and demonstrate a novel protocol for detecting spin–charge separation through an off-diagonal dissipative response. Our key finding reveals that when one spin component is subjected to dissipation, the other exhibits a universal temporal crossover following a $t^3$-to-$t$ law. This distinct dynamical signature—whose coefficients are governed by the different anomalous dimensions and velocities of spinons and holons—offers direct and unambiguous evidence of spin–charge separation. Our theoretical framework, developed via bosonization, is corroborated by large-scale tDMRG simulations. This work establishes off-diagonal dissipation as a powerful and general probe for quantum fractionalization in synthetic quantum matter. Moreover, the interplay between dissipation and non-equilibrium criticality in such settings constitutes a rich and largely uncharted frontier.

\emph{Acknowledgment}. --- This work is supported by the National Natural Science Foundation of China (Grants No. 12174358), the National Key R\&D Program of China (Grant No. 2022YFA1405302), and NSAF (Grant No. U2330401).

\end{document}